# Spin-neutral currents for spintronics

Ding-Fu Shao,[1,*] Shu-Hui Zhang,[2] Ming Li,[1] Chang-Beom Eom[3] and Evgeny Y. Tsymbal[1,†]

[1] *Department of Physics and Astronomy & Nebraska Center for Materials and Nanoscience,
University of Nebraska, Lincoln, Nebraska 68588-0299, USA*

[2] *College of Mathematics and Physics, Beijing University of Chemical Technology
Beijing 100029, People's Republic of China*

[3] *Department of Materials Science and Engineering, University of Wisconsin-Madison, Madison, WI 3706, USA*

**Electric currents carrying a net spin polarization are widely used in spintronics, whereas globally spin-neutral currents are expected to play no role in spin-dependent phenomena. Here we show that, in contrast to this common expectation, spin-independent conductance in compensated antiferromagnets and normal metals can be efficiently exploited in spintronics, provided their magnetic space group symmetry supports a non-spin-degenerate Fermi surface. Due to their momentum-dependent spin polarization, such antiferromagnets can be used as active elements in antiferromagnetic tunnel junctions (AFMTJs) and produce a giant tunneling magnetoresistance (TMR) effect. Using $RuO_2$ as a representative compensated antiferromagnet exhibiting spin-independent conductance along the [001] direction but a non-spin-degenerate Fermi surface, we design a $RuO_2/TiO_2/RuO_2$ (001) AFMTJ, where a globally spin-neutral charge current is controlled by the relative orientation of the Néel vectors of the two $RuO_2$ electrodes, resulting in the TMR effect as large as ~500%. These results are expanded to normal metals which can be used as a counter electrode in AFMTJs with a single antiferromagnetic layer or other elements in spintronic devices. Our work uncovers an unexplored potential of the materials with no global spin polarization for utilizing them in spintronics.**

The field of spintronics utilizes the spin degree of freedom in condensed matter for information processing and storage[1]. Most spintronic applications rely on electric currents with sizable spin polarization for detection or manipulation of the magnetic order parameter in spintronic devices. A typical and widely used spintronic device is the magnetic tunnel junction (MTJ), where a longitudinal charge current spin polarized by one ferromagnetic metal quantum-mechanically tunnels into another ferromagnetic metal through an insulating barrier layer[2,3]. Conductance of the MTJ is controlled by the relative magnetization orientation of the two ferromagnetic electrodes, resulting in a tunneling magnetoresistance (TMR) effect[4].

Contrary to the spin-polarized currents, spin-neutral currents are usually considered impractical for spintronics due to being unable to directly interact with the magnetic order parameter. This fact challenges spintronics based on non-ferromagnetic materials, such as compensated antiferromagnets, which normally do not support spin-polarized currents. Due to being robust against magnetic perturbations, the absence of stray fields, and ultrafast spin dynamics, antiferromagnets are considered as outstanding candidates to replace the widely used ferromagnets in the next generation spintronics[5-9]. This promising route has been recently stimulated by the demonstrated control of the antiferromagnetic Néel vector by spin-orbit torques[10,11]. However, the absence of a net magnetization and hence spin-independent conductance makes the electrical detection of the Néel vector using conventional methods, such as TMR measurements, unfeasible[5]. So far, the electrical detection of the Néel vector has been performed using anisotropic[10,11] or spin-Hall[12-15] magnetoresistance. Unfortunately, both methods suffer from relatively small signals easily influenced by perturbations[16] and require multiple in-plane terminals resulting in large device dimensions[7]. Antiferromagnetic spin valves[17-20] and antiferromagnetic tunnel junctions (AFMTJs)[21,22] have been theoretically proposed, promising, in some cases, sizable magnetoresistance effects. However, these magnetoresistance effects rely on perfect interfaces and switching the interfacial magnetic moment alignment between parallel and antiparallel. This mechanism is not robust against disorder and interface roughness inevitable in experimental conditions. Recent efforts have been aimed at exploring unconventional methods for the Néel vector detection based on topological properties[9,23-25] but require an experimental confirmation.

One promising direction is to create spin-polarized currents in antiferromagnets. Recently, it has been predicted that certain types of compensated antiferromagnets exhibit a momentum-dependent spin splitting of the Fermi surface[26-28], resulting in spin-polarized currents along certain crystallographic orientations[29-31]. These predictions indicate that these antiferromagnets can work as ferromagnets in spintronic devices, which broadens the range of materials useful for spintronics.

Here, we embark on a different path and argue that *globally spin-independent* conductance in compensated antiferromagnets can be efficiently used in spintronics, provided their crystal symmetry supports a non-spin-degenerate Fermi surface and thus momentum-dependent spin polarization. While such a spin polarization is cancelled out in the net conductance due to being antisymmetric with respect to certain symmetry operations, its presence in the momentum space can be functionalized if such an antiferromagnet is combined with another similar antiferromagnet in a spintronic device such as an AFMTJ. In this case, the resistance change of the AFMTJ occurs in response to the orientation of the antiferromagnetic Néel vector due to changing matching conditions between the spin-polarized conduction channels in the two metal electrodes. These considerations can be expanded to normal (nonmagnetic) metals with spin-orbit coupling where the combined space inversion-time reversal symmetry is broken, indicating that they can also be utilized in spintronics despite globally spin-neutral currents.

## Results

**Spin polarized conduction channels.** To explore the possible use of spin-neutral currents in a spintronic device, we first consider ballistic conductance of a material under investigation. Since the ballistic conductance is determined by the number of conduction channels, i.e. propagating Bloch states at the Fermi energy, it can provide an important characteristic of a spintronic device where this material is used as a metal electrode[32,33]. In the absence of spin-orbit coupling, the ballistic conductance $g$ per unit area along the $z$ direction can be obtained in terms of two spin components as follows[34]:

$$g = g^\uparrow + g^\downarrow = \frac{e^2}{h}\sum_{\vec{k}_\parallel}\left(N_\parallel^\uparrow + N_\parallel^\downarrow\right), \quad (1)$$

$$N_\parallel^\sigma(\vec{k}_\parallel) = \frac{\hbar}{2}\sum_n \int |v_{nz}^\sigma|\frac{\partial f}{\partial E_n^\sigma(\vec{k})} dk_z. \quad (2)$$

Here $\sigma$ denotes the spin component ↑ or ↓, $\vec{k}$ is the wave vector in the three-dimensional Brillouin zone, $N_\parallel^\sigma(\vec{k}_\parallel)$ is the number of conduction channels (integer) at the transverse wave vector $\vec{k}_\parallel = (k_x, k_y)$ for spin $\sigma$, $E_n^\sigma$ is energy for the $n$-th band, $v_{nz}^\sigma = \frac{\partial E_n^\sigma(\vec{k})}{\hbar \partial k_z}$ is the band velocity along the transport $z$ direction, and $f$ is the Fermi distribution function.

The net transport spin polarization is defined by

$$p = \frac{g^\uparrow - g^\downarrow}{g} \quad (3)$$

and represents an important quantity useful in spintronics. For example, in a crude approximation of $\vec{k}_\parallel$-independent transmission between two ferromagnetic electrodes with spin polarizations $p_1$ and $p_2$ in an MTJ, the TMR effect is given by the well-known Julliere's formula[2] $TMR = \frac{2p_1 p_2}{1 - p_1 p_2}$. Clearly, a larger spin polarization of the electrodes favors a larger TMR.

A large spin polarization $p$ is generally expected for ferromagnets where the finite net magnetization breaks time reversal symmetry $\hat{T}$. The latter flips the spin $\sigma$ and changes sign of $\vec{k}_\parallel$, resulting in $\hat{T} N_\parallel^\uparrow(\vec{k}_\parallel) = N_\parallel^\downarrow(-\vec{k}_\parallel)$. Compensated antiferromagnets do not have net magnetization and hence (with some exceptions[29-31]) do not support the macroscopic spin-polarized current. However, even though macroscopically the net transport spin polarization $p$ is absent, microscopically the conductance could be spin polarized as reflected in the spin polarization of conduction channels at $\vec{k}_\parallel$:

$$p_\parallel(\vec{k}_\parallel) = \frac{N_\parallel^\uparrow - N_\parallel^\downarrow}{N_\parallel^\uparrow + N_\parallel^\downarrow}. \quad (4)$$

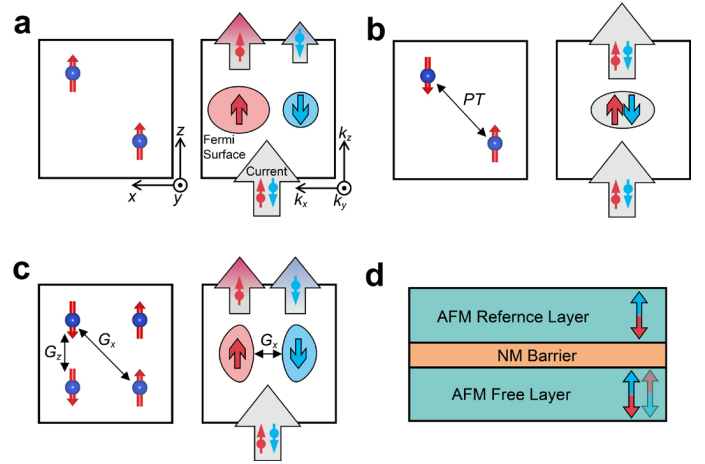

**Fig. 1 Spin-polarized conduction channels in different types of magnetic materials. a**, Schematics of the atomic structure (left) and spin-polarized Fermi surface (right) for a ferromagnet. A spin-neutral current passing through the ferromagnet becomes spin polarized. **b**, Schematics of the atomic structure (left) and spin-degenerate Fermi surface (right) for a compensated antiferromagnet where the net magnetization is forbidden by $\hat{P}\hat{T}$ symmetry. A spin-neutral current passing through this antiferromagnet remains spin neutral. **c**, Schematics of the atomic structure (left) and non-spin-degenerate Fermi surface congruent for opposite spins (right) for a compensated antiferromagnet where the net magnetization is forbidden by glide symmetries $\hat{G}_x$ and $\hat{G}_z$. A spin-neutral current passing through this antiferromagnet remains globally spin neutral but has a momentum-dependent spin polarization. **d**, Schematics of an AFMTJ where two antiferromagnetic (AFM) layers are separated by a nonmagnetic (NM) barrier layer. The Néel vector (indicated by arrows) of the bottom free layer can be switched resulting in the TMR effect.

If both electrodes in a two-terminal spintronic device are made of materials with zero net spin polarization but have spin-polarized conduction channels, this momentum-dependent spin polarization will be reflected in the device conductance and can be functionalized through the antiferromagnetic Néel vector. Indeed, in the transport regime conserving spin (no spin-orbit coupling) and wave vector $\vec{k}_\parallel$ (no diffuse scattering), the device conductance is largely affected by the spin matching of the conduction channels $\vec{k}_\parallel$ of the electrodes. If their spin polarization changes in response to the Néel vector rotation in an antiferromagnetic electrode, this alters the net conductance of the device. Thus, although the spin polarization of the charge current remains zero, the device conductance reflects the $\vec{k}_\parallel$- dependent spin polarization of the conduction channels.

Next, we identify magnetic space group symmetry requirements for crystals to exhibit the $\vec{k}_\parallel$-dependent spin polarization. Obviously, in ferromagnets, the conduction channels are spin polarized due to the spin-dependent Fermi surface. Thus, passing a spin-neutral current through a ferromagnetic material makes it spin polarized (Fig. 1a). On the contrary, most compensated antiferromagnets contain symmetries that not only prevent the net magnetization but also lead to a spin-degenerate Fermi surface and thus spin-independent conduction channels. For example, if a compensated antiferromagnet exhibits $\hat{P}\hat{T}$ symmetry, where $\hat{P}$ and $\hat{T}$ are space inversion and time reversal symmetries, respectively, $p_\parallel = 0$ due to $\hat{P}\hat{T}N_\parallel^\uparrow(\vec{k}_\parallel) = N_\parallel^\downarrow(\vec{k}_\parallel)$. This property follows from the spin-degenerate Fermi surface due to $\hat{P}\hat{T}E_n^\uparrow(\vec{k})= E_n^\downarrow(\vec{k})$ (Fig. 1b). The spin degeneracy also appears in compensated antiferromagnets with $\hat{T}\hat{t}$ symmetry ($\hat{t}$ is half a unit cell translation) in the absence of spin-orbit coupling.

The spin degeneracy is however broken in compensated antiferromagnets belonging to magnetic space groups with violated $\hat{P}\hat{T}$ and $\hat{T}\hat{t}$ symmetries[26]. The vanishing net magnetization in such antiferromagnets originates from the combination of some other magnetic space group symmetries of the crystal. For example, Fig. 1c shows a collinear antiferromagnet with the Néel vector pointing along the $z$ direction. The zero net magnetization in this antiferromagnet is guaranteed by two glide symmetries $\hat{G}_x$ and $\hat{G}_z$, where $\hat{G}_l = \{\hat{M}_l|\hat{t}\}$ represents mirror symmetry $\hat{M}_l$ with a mirror plane normal to vector $\vec{l}$ combined with translation $\hat{t}$. The symmetry transformation $\hat{G}_x N_\parallel^\uparrow(k_x, k_y) = N_\parallel^\downarrow(-k_x, k_y)$ flips the spin and thus according to Eqs. (1) and (2) results in a vanishing net spin polarization $p$ for the current along the $z$ direction. On the other hand, due to the Néel vector pointing along the $z$ axis, the symmetry transformation $\hat{G}_z E_n^\sigma(\vec{k}_\parallel, k_z) = E_n^\sigma(\vec{k}_\parallel, -k_z)$ conserves the spin $\sigma = \uparrow, \downarrow$ of the conduction modes at $\vec{k}_\parallel$ and there is no symmetry operation which would enforce $p_\parallel = 0$. The presence of spin-polarized conduction channels in this type of antiferromagnets can be understood in terms of two congruent (but not identical) up- and down-spin Fermi surfaces, which are transformed to each other by the symmetry transformation $\hat{G}_x$ (Fig. 1c). In this case, each conduction channel is spin polarized (except high-symmetry $\vec{k}_\parallel$ points invariant to $\hat{G}_x$), whereas the net conductance is spin neutral.

Due to the non-spin-degenerate Fermi surface and spin-polarized conduction channels, the globally spin-neutral conduction of the compensated antiferromagnets can be exploited in spintronic devices, such as AFMTJs. Figure 1d shows an AFMTJ which contains two identical antiferromagnetic electrodes separated by a nonmagnetic insulating spacer. The antiferromagnets are assumed to have spin-polarized conduction channels along the out-of-plane transport direction. The functionality of the AFMTJ is controlled by the relative orientation of the Néel vector of the two antiferromagnetic electrodes. In the parallel state, the spin-polarized conduction channels of the electrodes perfectly match, resulting in a low resistance state. In the antiparallel state, the spin polarized conduction channels are mismatched, resulting in a high resistance state.

**Electronic structure of RuO$_2$.** To demonstrate this spintronic functionality, we consider the recently discovered room-temperature antiferromagnetic metal RuO$_2$ [35] suitable for realizing the proposed AFMTJ. RuO$_2$ exhibits interesting properties such as spin splitting without spin-orbit coupling[36], a crystal Hall effect[37], and a magnetic spin Hall effect[31,38]. RuO$_2$ has a rutile structure with an out-of-plane Néel vector (Fig. 2a) and magnetic space group $P4_2'/mnm'$, which contains glide $\hat{G}_x = \{\hat{M}_x|(\frac{1}{2},\frac{1}{2},\frac{1}{2})\}$, $\hat{G}_y = \{\hat{M}_y|(\frac{1}{2},\frac{1}{2},\frac{1}{2})\}$ and mirror $\hat{M}_z$ symmetries. In the absence of spin-orbit coupling, the energy bands are spin degenerate at the $k$-planes invariant to $\hat{G}_x$ and $\hat{G}_y$, such as $k_x = 0, \frac{\pi}{2}$ or $k_y = 0, \frac{\pi}{2}$. This is evident from our first-principles density functional theory (DFT) calculations. As seen from Fig. 2b, the energy bands of RuO$_2$ are spin degenerate along the Γ-X, Γ-Z, X-M, Z-R, and R-A directions lying in these glide-invariant planes. On the other hand, a large spin splitting appears along the directions away from these planes, such as Γ-M and Z-A.

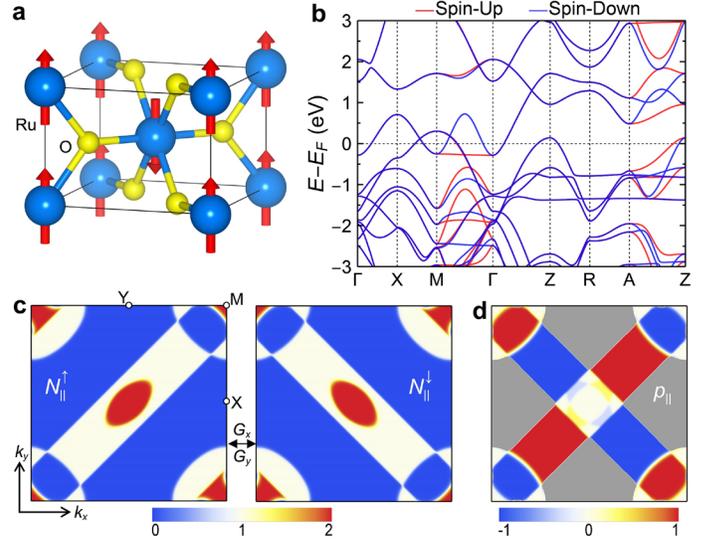

**Fig. 2 Electronic properties of RuO$_2$. a**, The atomic and magnetic structures of RuO$_2$. **b**, The calculated band structure of RuO$_2$. **c**, The number of $\vec{k}_\parallel$-resolved conduction channels in the 2D Brillouin zone of RuO$_2$ for spin up $N_\parallel^\uparrow$ (left) and spin down $N_\parallel^\downarrow$ (right). High-symmetry $\vec{k}_\parallel$-points are indicated. $N_\parallel^\uparrow$ and $N_\parallel^\downarrow$ can be transformed to each other by the glide transformations $\hat{G}_x$ or $\hat{G}_y$. **d**, Spin polarization of conduction channels $p_\parallel(\vec{k}_\parallel)$. Grey contrast indicates regions where $N_\parallel^\uparrow = N_\parallel^\downarrow = 0$ and thus $p_\parallel$ undefined.

The spin-dependent band structure leads to the momentum-dependent spin polarization of the conduction channels along the $z$ direction. We explicitly demonstrate this by calculating the number of conduction channels $N_\parallel^\sigma(\vec{k}_\parallel)$ in the two-dimensional (2D) Brillouin zone of $RuO_2$. As seen from Fig. 2c, the distributions of $N_\parallel^\uparrow$ and $N_\parallel^\downarrow$ in the $(k_x, k_y)$ plane have congruent shapes and are symmetric with respect to the $\hat{G}_x$ and $\hat{G}_y$ symmetry transformations, which enforce a zero global spin polarization in the conductance along the $z$ direction. On the contrary, as seen from Fig. 2d, the $\vec{k}_\parallel$-dependent spin polarization $p_\parallel$ remains finite across a sizable portion of the 2D Brillouin zone. This demonstrates that $RuO_2$ (001) exhibits a globally spin-neutral conductance through spin-polarized conduction channels.

**TMR in a $RuO_2/TiO_2/RuO_2$ AFMTJ.** Next, we design an AFMTJ using $RuO_2$ (001) as electrodes and $TiO_2$ (001) as an insulating barrier layer. Due to both having a rutile structure and a similar lattice constant, this AFMTJ is feasible in practice. Figure 3a shows the atomic structure of the $RuO_2/TiO_2/RuO_2$ (001) supercell, which is used in our DFT and quantum transport calculations and includes 8 $TiO_2$ layers in the center and 10 $RuO_2$ layers on each side. We find that a wide band gap of $TiO_2$ is well maintained in this heterostructure, and the Fermi energy $E_F$ is located deeply inside the band gap (Fig. 3b).

The $RuO_2/TiO_2/RuO_2$ (001) structure in Fig. 3a is then used as the scattering region of the AFMTJ connected to two semi-infinite $RuO_2$ (001) electrodes for calculating transmission. The transmission is obtained for parallel (Fig. 4a) and antiparallel (Fig. 4b) alignments of the Néel vectors of the electrodes. For the parallel-aligned AFMTJ, the $\vec{k}_\parallel$-resolved transmission $T_P^\sigma(\vec{k}_\parallel)$ is shown in Fig. 4c for spin up (left panel) and spin down (right panel). The clearly seen spin asymmetry between $T_P^\uparrow(\vec{k}_\parallel)$ and $T_P^\downarrow(\vec{k}_\parallel)$ reflects the related asymmetry in the distribution of the spin-polarized conduction channels in $RuO_2$ (Fig. 2c). The suppressed transmission near the Brillouin zone corners is due to a larger decay rate of the evanescent states for $\vec{k}_\parallel$ away from the zone center.

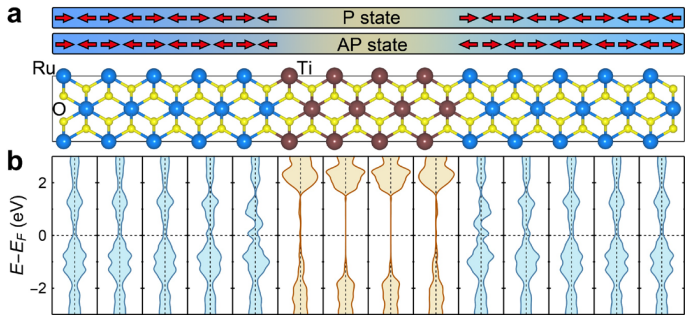

**Fig. 3 Atomic and electronic structure of a $RuO_2/TiO_2/RuO_2$ supercell. a,b,** The atomic structure (**a**) and layer resolved density of states (DOS) (**b**) of the $RuO_2/TiO_2/RuO_2$ supercell. Each panel in **b** contains two atomic layers of $MO_2$ (M = Ru, Ti) and has left and right subpanels corresponding to up- and down-spin states, respectively. The arrows in (**a**) indicate the magnetic moments of Ru atoms.

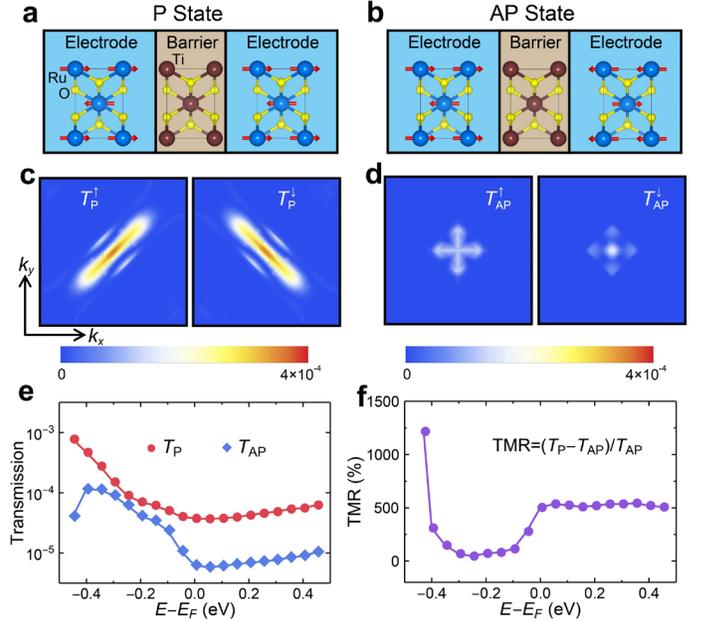

**Fig. 4 Giant TMR in $RuO_2/TiO_2/RuO_2$ AFMTJ. a, b,** The atomic and magnetic structures of $RuO_2/TiO_2/RuO_2$ AFMTJ for parallel (P) (**a**) and antiparallel (AP) (**b**) alignment of the Néel vectors. **c, d,** The calculated $\vec{k}_\parallel$-resolved transmission in the 2D Brillouin zone for the AFMTJ in P (**c**) and AP (**d**) states. **e,** Total transmission as a function of energy for the AFMTJ in P (red dots) and AP (blue dots) states. **f,** TMR as a function of energy.

For the antiparallel-aligned AFMTJ, the transmission $T_{AP}^\sigma(\vec{k}_\parallel)$ is blocked for the wave vectors $\vec{k}_\parallel$ with no conduction channels in one of the spin states, i.e. $N_\parallel^\sigma(\vec{k}_\parallel) = 0$ for $\sigma = \uparrow$ or $\sigma = \downarrow$. These are the regions where $p_\parallel = \pm 1$ or $N_\parallel^\uparrow = N_\parallel^\downarrow = 0$ in Figs. 2c and 2d. As a result, only the states near the zone center where the spin channels are degenerate (enforced by $\hat{G}_x$ and $\hat{G}_y$) contribute to the transmission (Fig. 4d). This leads to the total transmission being much smaller for the antiparallel state ($T_{AP}$) than for the parallel state ($T_P$) (Fig. 4e). At the Fermi energy $E_F$, we find the TMR ratio $(T_P - T_{AP})/T_{AP}$ as large as ~500%. This value is comparable to the values obtained for the well-known Fe/MgO/Fe (001) MTJs [39,40] which are currently used in magnetic random-access memories. The giant TMR appears not only for $E = E_F$ but also for the energies around the Fermi level, with the smallest value of ~50% at $E = E_F - 0.25$ eV (Fig. 4f).

This fact indicates that the large TMR will be sustained under an applied bias voltage.

The predicted TMR is largely independent of the interface terminations and the relative alignment of the interface magnetic moments, as follows from our explicit DFT calculations (Supplementary Figs. S1 and S2). This distinguishes our results from the previous findings[21,22], where the interface termination controls TMR and implies that the predicted TMR effect is likely less sensitive to the interface roughness than that in the previous studies. The bulk origin of TMR in the proposed AFMTJs makes it also more robust against other types of disorder, as long as the crystallinity of the tunnel junction and the direct tunneling transport mechanism are maintained (see Section C of Supplemental Material).

**Discussion**

The above properties are sustained in the presence of spin-orbit coupling. This is due to the momentum-dependent spin polarization in rutile antiferromagnets being inherited from the antiferromagnetic order rather than spin-orbit coupling[27, 31, 37]. In RuO$_2$, the $\hat{M}_z$ symmetry transformation reverses the wave vector component $k_z$, conserves the spin component $\sigma_z$ but flips $\sigma_x$ and $\sigma_y$. As a result, the conduction modes at $\vec{k}_\parallel$ are spin polarized purely along the $z$ axis and the $x$- and $y$-spin components vanish (see Supplementary Section D). Supplementary Fig. S5e shows the results of the DFT calculation for RuO$_2$ in the presence of spin-orbit coupling. It is seen that the spin polarization of the most conduction channels is well preserved, indicating that the giant TMR is robust against spin-orbit coupling in the RuO$_2$ based AFMTJ.

The spin-polarized conduction channels are not limited to RuO$_2$, but typical for a wide group of materials with violated $\hat{P}\hat{T}$ symmetry, including those with a noncollinear antiferromagnetic order[29]. For the antiferromagnets with magnetization being compensated by combined mirror and/or rotation symmetries, the spin-polarized conduction channels can exist purely due to the antiferromagnetic order, even in the absence of spin-orbit coupling[26-31]. On the other hand, in compensated antiferromagnets with $\hat{T}\hat{t}$ symmetry, the spin-polarized conduction channels appear due to the spin degeneracy lifted by spin-orbit coupling. Such antiferromagnets can also be used in AFMTJs if they have sizable spin-orbit splitting.

These considerations can be expanded to normal metals with broken space inversion symmetry, where the band spin degeneracy is lifted by spin-orbit interaction. In these materials, the two conduction channels with opposite spin polarizations are linked by the time reversal symmetry operation. For example, in topological metal TaN[41, 42], the conduction channels along the [001] direction carry the spin polarization pointing along the same [001] direction (Supplementary Fig. S6). This property is enforced by the $\hat{M}_z$ mirror symmetry. While the momentum-dependent spin polarization in non-centrosymmetric normal metals is fixed by their crystal symmetry and band structure, they can be used in spintronics in conjunction with antiferromagnets. For example, the antiferromagnetic reference layer in the AFMTJ in Fig. 1d can be replaced by a normal metal layer. Alternatively, one can create an antiferromagnet/normal metal interface. In such systems with a single antiferromagnetic layer, the spintronic functionality is controlled by the Néel vector orientation that regulates a matching of the conduction channels in the antiferromagnetic and normal metal layers.

Note that in junctions with a single antiferromagnetic layer, reversal of the Néel vector is equivalent to the time-reversal transformation which does not change the resistance. However, the resistance changes with rotation of the Néel vector, resulting in a tunnelling anisotropic magnetoresistance (TAMR) effect[43].

Another possibility is to utilize non-centrosymmetric insulators as a tunneling barrier layer in an AFMTJ. Due to the broken space inversion symmetry and spin-orbit coupling, the evanescent states in these insulators are spin-polarized[44, 45]. Therefore, the Néel vector of the free antiferromagnetic layer can be used to control the matching between the propagating Bloch states in the antiferromagnetic electrode and the evanescent gap states in the barrier resulting in a TAMR effect. An additional useful functionality of this kind of tunnel junctions may be provided by a switchable polarization of the non-centrosymmetric insulating barrier layer if it is ferroelectric[45].

The proposed use of spin-neutral currents in spintronics is feasible from the experimental perspective. For example, the proposed RuO$_2$/TiO$_2$/RuO$_2$ (001) AFMTJ has all rutile structure with a good match of the RuO$_2$ and TiO$_2$ lattice constants and thus can be grown epitaxially preserving crystallinity of the overall heterostructure. The Néel vector of the antiferromagnetic free layer can be switched by a spin-orbit torque via the spin current from an adjacent heavy metal layer generated by an in-plane charge current[8,46]. With the in-plane writing path and out-of-plane reading path, only two in-plane terminals and one out-of-plane terminals are required for such an AFMTJ, which is desirable for nanoscale spintronic applications. In addition, the large magnitude of TMR indicates a possibility of a strong spin transfer torque in the AFMTJs, which may be robust against disorder[47] and may offer an alternative way to switch the Néel vector.

In conclusion, we have proposed that globally spin-neutral currents flowing through oppositely spin-polarized conduction channels can be efficiently used in spintronics. Such currents exist in compensated antiferromagnets and normal metals with the magnetic space group symmetries which lift the spin-degeneracy of the Fermi surface. In the heterostructures, such as antiferromagnetic tunnel junctions or antiferromagnet/normal metal interfaces, these currents can be controlled by the Néel vector orientation providing a useful functionality for spintronics. Based on first-principles density functional theory combined with quantum transport calculations, we have demonstrated such functionality using a room-temperature antiferromagnetic metal $RuO_2$ as electrodes in a $RuO_2/TiO_2/RuO_2$ AFMTJ and predicted a giant TMR effect of ~500%. Our work uncovers an unexplored potential of the materials with no global spin polarization for utilizing them in spintronics. We hope therefore that our predictions will stimulate experimental investigations of these materials and the associated phenomena.

*Note added*: During the review of this manuscript, we became aware of the relevant work by Šmejkal *et al.* posted recently [48].

## Methods

The atomic and electronic structures shown in Figs. 2a,b, 3, S5a, and S6a,b of the systems are calculated using the projector augmented wave (PAW) method[49] implemented in the VASP code[50]. A plane-wave cut-off energy of 500 eV and a 16 × 16 × 16 $\vec{k}$-point mesh in the irreducible Brillouin zone are used in the calculations. The exchange and correlation effects are treated within the generalized gradient approximation (GGA) developed by Perdew-Burke-Ernzerhof (PBE) [51]. The GGA+U functional[52,53] with $U_{eff}$ = 2 eV on Ru 4$d$ orbitals and $U_{eff}$ = 5 eV on Ti 3$d$ orbitals is included in the calculations.

The transport properties shown in Figs. 4 and S1-S3 are calculated using the non-equilibrium Green's function formalism (DFT+NEGF approach)[54,55], as implemented the Atomistic Simulation Toolkit (ATK) distributed in the QuantumWise package (Version 2015.1) [56,57]. The atomic structures are relaxed by VASP and the nonrelativistic Fritz-Haber-Institute (FHI) pseudopotentials using a single-zeta-polarized basis. The spin polarized GGA+U functional[51,52] with $U_{eff}$ = 2.3 eV on Ru 4$d$ orbitals and $U_{eff}$ = 5 eV on Ti 3$d$ orbitals is included in the calculations. A cut-off energy of 75 Ry and a 11×11×101 $\vec{k}$-point mesh are used for the self-consistent calculations to eliminate the mismatch of the Fermi level between the electrodes and the central region. Unless mentioned in the text, the transmission is calculated using an adaptive $\vec{k}$-point mesh. These parameters are confirmed to yield a good balance between the computational time and accuracy.

The tight-binding Hamiltonians of $RuO_2$ and TaN are obtained using Wannier90 code [58] utilizing the maximally localized Wannier functions[59]. A 500 × 500 × 500 $\vec{k}$-point mesh and the adaptive smearing method[60] are used to calculate the $\vec{k}_{\parallel}$-resolved ballistic conductance shown in Figs. 2c,d, S5e and S6c,d. The spin-projected Fermi surfaces of $RuO_2$ with spin-orbit coupling shown in Figs. S5b-d are calculated using WannierBerri code[61,62].

Figures are plotted using VESTA [63], FermiSurfer [64], gnuplot[65], and the SciDraw scientific figure preparation system[66].

## Data availability

The data that support the findings of this study are available from the corresponding author upon reasonable request.

## Acknowledgments

The authors thank Bo Li for helpful discussions. This work was supported by the Vannevar Bush Faculty Fellowship (ONR grant N00014-20-1-2844) and by the National Science Foundation (NSF) through the MRSEC (NSF Award DMR-1420645) and EPSCoR RII Track-1 (NSF Award OIA-2044049) programs. S.-H. Z. thanks the support of National Science Foundation of China (NSFC Grant No. 12174019). Computations were performed at the University of Nebraska Holland Computing Center.

## Author contributions

D.-F.S. and E.Y.T. conceived this project. D.-F.S. performed symmetry analysis and calculated electronic properties for $RuO_2$ and TaN. S.H.Z calculated the ballistic conductance of $RuO_2$ and TaN. D.-F.S and M.L. calculated the transport properties of $RuO_2/TiO_2/RuO_2$ AFMTJ. D.F.S, S.H.Z, M.L. and E.Y.T. analyzed the results and discussed with C.B.E. D.-F.S. and E.Y.T. wrote the manuscript. All authors contributed to the final version of the manuscript.

## Competing interests

The authors declare no competing interests.

* dfshao@unl.edu
† tsymbal@unl.edu